# A Detailed Dynamical Investigation of the Proposed QS Virginis Planetary System

J. Horner[1], R. A. Wittenmyer[1], T. C. Hinse[2], J. P. Marshall[3], A. J. Mustill[3] & C. G. Tinney[1]

[1] Department of Astrophysics and Optics, School of Physics, University of New South Wales, Sydney 2052, Australia
[2] Korea Astronomy and Space Science Institute, 776 Daedeokdae-ro Yuseong-gu 305-348 Daejeon, Korea
[3] Departamento de Física Teórica, Facultad de Ciencias, Universidad Autónoma de Madrid, Cantoblanco, 28049 Madrid, Spain

**Abstract**
In recent years, a number of planetary systems have been proposed to orbit evolved binary star systems. The presence of planets is invoked to explain observed variations in the timing of mutual eclipses between the primary and secondary components of the binary star system. The planets recently proposed orbiting the cataclysmic variable system QS Virginis are the latest in this on-going series of "extreme planets".

The two planets proposed to orbit QS Virginis would move on mutually crossing orbits – a situation that is almost invariably unstable on very short timescales. In this work, we present the results of a detailed dynamical study of the orbital evolution of the two proposed planets, revealing that they are dynamically unstable on timescales of less than one thousand years across the entire range of orbital elements that provide a plausible fit to the observational data, and regardless of their mutual orbital inclination. We conclude that the proposed planets around the cataclysmic variable QS Virginis simply cannot exist.

## 1. Introduction

Over the last few years, the number of exoplanets that have been announced has risen dramatically. There are two main reasons for the rapid explosion in the number of known planets around other stars. First, the Kepler spacecraft (e.g. Borucki et al., 2010, 2011) has proven hugely successful, finding (to date) 132 confirmed exoplanets, and a further 2740 planet candidates[1]. Secondly, the various radial velocity programs being carried out worldwide (e.g. HARPS, Mayor et al., 2003, Udry et al., 2007; AAPS, Tinney et al., 2001, Wittenmyer et al., 2012b; the Texas Planet Search, Cochran et al., 2004; Wittenmyer et al., 2009; Robertson et al., 2012a, b; the California planet survey, Howard et al., 2010; and the Lick-Carnegie planet search, Haghighipour et al., 2010) have been able to both probe to lower radial velocities (enabling the detection of lower mass planets) and have access to results spanning a longer observational baseline (enabling the detection of planets of ever longer orbital period). The discoveries announced using these two premier methods of exoplanet detection are typically considered to be robust, and researchers take great pains to rule out other explanations for the observations that infer the presence of planets, prior to announcing their discoveries (e.g. Tinney et al., 2011; Robertson et al., 2013).

In recent years, a dozen post-common-envelope binary stars have been touted as planet hosts, on the basis of observed variations in the timing of eclipses between the two binary components (Zorotovic & Schrieber 2013). Were the components of the binaries isolated in space, one might expect their eclipses to occur perfectly periodically (aside from a small and predictable variation resulting from relativistic effects e.g. Meliani et al., 2000), as one component passes in front of the other along our line of sight to the system. Any variation from that perfect periodicity must be the result of some other physical process – either a non-gravitational interaction between the two stars (such as the Applegate mechanism, Applegate, 1992) or the gravitational influence of one or more unseen companions. The presence of unseen companions would cause the two central stars to rock back and forth along our line of sight, as they moved around the centre of mass of the whole system. For this reason, the light from their eclipses would sometimes arrive a little early at the Earth (when the stars are slightly closer to us), and sometimes slightly late (when they are further away).

---

[1] Numbers taken from the Kepler home page, http://kepler.nasa.gov/, on the 17th May, 2013. The process by which a planet moves from the candidate list to being confirmed requires a significant amount of ground-based follow-up work – one of the downsides of the fact that many of Kepler's target stars are faint and distant objects. In the coming years, it is likely that the great majority of the candidate planets will eventually be confirmed (Lissauer et al., 2012).

Based on observed variations in the timing of eclipses, unseen companions have been announced orbiting a number of such binaries (HW Virginis, Lee et al., 2009; NN Serpentis, Beuermann et al., 2010, 2013; RZ Draconis, Yang et al., 2010; HU Aquarii, Qian et al., 2011; SZ Herculis, Lee et al., 2012; NSVS 14256825, Almeida, Jablonski & Rodrigues, 2013). The orbits of the proposed companions in these systems are based on a purely Keplerian fit to the observed data, which takes no account of any potential interactions between the objects in question. It is therefore important to consider whether the companions proposed to explain the observed variations in eclipse timing are dynamically feasible.

We have previously examined the dynamical evolution of the proposed companions in each of these systems. In the cases of HU Aquarii (Horner et al., 2011; Wittenmyer et al., 2012a), HW Virginis (Horner et al., 2012a), NSVS 14256825 (Wittenmyer et al., 2013a), SZ Herculis (Hinse et al., 2012) and RZ Draconis (Hinse et al., 2013a), we found that the proposed companions were dynamically unstable on timescales of just a few thousand years (or less). In other words, the observed variations in timing of eclipses between the components of the binary systems in question must have some other explanation. In the case of NN Serpentis, the proposed planets do stand up to dynamical scrutiny (Horner et al., 2012b) – with broad regions of dynamical stability encompassed within the ±3 sigma uncertainties on the orbital solution proposed in the discovery work (Beuermann et al., 2010). However, recent work (Mustill et al., 2013) has shown that, while the planets proposed in that system move on orbits that would be dynamically stable at the current epoch, it is almost certain that the observed signal is not attributable to planets that formed around the youthful binary (in a proto-planetary disc), then survived the binary's post-main sequence evolution whilst moving to their current orbits. Once again, a planetary explanation for the proposed planets does not stand up to close scrutiny. Interestingly, recent work (Hinse et al., 2013b) suggests that, at least in the case of NSVS 14256825, the uncertainties on the eclipse timings may have been underestimated. Increasing the uncertainty in the observed times of mid-eclipse to ±5 seconds removes all evidence for statistically significant periodic eclipse timing variations, and the data can be fitted simply by the linear ephemeris that one would expect for an isolated pair of stars in orbit around their common centre of gravity.

In this work, we consider the recently announced companions to the cataclysmic variable binary system QS Virginis (Almeida & Jablonski, 2011). In section 2, we introduce the QS Virginis system, as described in that work, before describing our dynamical study of the system, and presenting our results, in section 3. Finally, in section 4, we present our conclusions.

## 2. The QS Virginis system.

The cataclysmic variable QS Virginis is a tightly bound binary star system, with an orbital period of 3 hours and 37 minutes (O'Donoghue et al., 2003). O'Donoghue et al. estimate that the primary star (a DA white dwarf) has a mass of $0.78 \pm 0.04$ M$_\odot$, whilst its companion (a dMe dwarf) has a mass of just $0.43 \pm 0.04$ M$_\odot$. Once per orbit, the primary is eclipsed by the secondary, with an eclipse duration of 14 minutes. For more details on the binary system itself, we direct the interested reader to O'Donoghue et al. (2003) and the more recent study of the accretion in the system, Matranga et al. (2012).

A number of authors have proposed the presence of an unseen companion in the QS Virginis system in order to explain observed variations in the timing of the eclipses between the two components. In the first such study, Qian et al. (2010) were able to fit the observed timing variations using a Keplerian term plus a quadratic trend. They suggested that the variations were best explained by the combination of the presence of an unseen companion, of mass around 6.4 times that of Jupiter, and an ongoing loss of angular momentum in the system through magnetic braking – a process which they claim is driving the evolution of the system from a hibernating cataclysmic variable (CV) state to becoming a more typical CV. Their proposed massive planetary companion had an orbital period of 7.86 years, and an orbital eccentricity of 0.37.

Parsons et al. (2010) added a number of new observations to the available dataset for QS Virginis, before once again attempting to explain the variations in observed eclipse timings by performing a Keplerian fit to the data.

They noted that they "… detect a large (~250 s) departure from linearity in the eclipse times of QS Vir which Applegate's mechanism fails to reproduce by an order of magnitude. The only mechanism able to drive this period change is a third body in a highly elliptical orbit". Their additional data-points, particularly those which can be seen at around cycle 35,000 in their Figure 10, force the fit to the observed data to be dramatically different to that proposed by Qian et al. (2010). Rather than a moderately eccentric Super-Jupiter on a 7.86 year orbit, their fit invokes the presence of a far more massive companion (0.05 $M_\odot$, or approximately 53 times Jupiter's mass), moving on a ~14 year orbit with an eccentricity of ~0.9. Because of the high eccentricity of the proposed companion, they discuss its plausibility and conclude that such a companion seems unlikely, but cannot be ruled out on the basis of our current understanding of the formation and evolution of such objects. We note that recent work (e.g. Wittenmyer et al., 2013b) has shown that radial velocity studies (which fit data in a very similar manner) sometimes find single, massive, eccentric companions which (upon the acquisition of further data) are sometimes revealed to instead be multiple companions moving on low eccentricity orbits, with significantly lower mass (e.g. Wittenmyer et al. 2012b). It therefore seems plausible that the eccentric companion proposed by Parsons et al. (2010) could be such a case, and that further observations could be used to show multiple companions in the QS Virginis system, moving on orbits with more reasonable eccentricity.

With that in mind, it is perhaps unsurprising that Almeida & Jablonski (2011) proposed that the QS Virginis system contains two unseen companions of masses 0.0086 and 0.054 times that of the Sun (respectively roughly 9 and 57 times the mass of Jupiter). However, their two-companion solution still features highly eccentric orbits and, in addition, places their massive companions on mutually crossing orbits, with periods that are remarkably close to one another. The orbital solution proposed in that work is shown in Table 1, with the nominal best fit orbits plotted in Figure 1.

|  | QS Vir (AB) b | QS Vir (AB) c |
|---|---|---|
| Orbital Period (years) | 14.40 ± 0.08 | 16.99 ± 0.07 |
| Semi-Major Axis[2] (AU) | 6.031 ± 0.051 | 7.043 ± 0.019 |
| Eccentricity | 0.62 ± 0.02 | 0.92 ± 0.02 |
| Argument of Periastron (°) | 180.0 ± 2.6 | 219 ± 3 |
| Mass[3] ($M_\odot$) | 0.0086 | 0.054 |

**Table 1:** The orbits of the two proposed companions to QS Virginis, as detailed in Almeida & Jablonski (2011). The orbits are highly eccentric (reminiscent of the orbits of comets in our own Solar system), and cross one another, as can be seen in Fig. 1.

---

[2] Calculated from the orbital period, with the binary as a 1.21 $M_\odot$ point mass; (O'Donoghue et al., 2003).
[3] The mass quoted here is the *minimum* mass for the planets (*m* sin *i*) – the mass derived assuming the companions orbit in the same plane as our line of sight. If the companion orbits are inclined to our line of sight by an angle *i*, then the true mass of the companions will be larger than this minimum value.

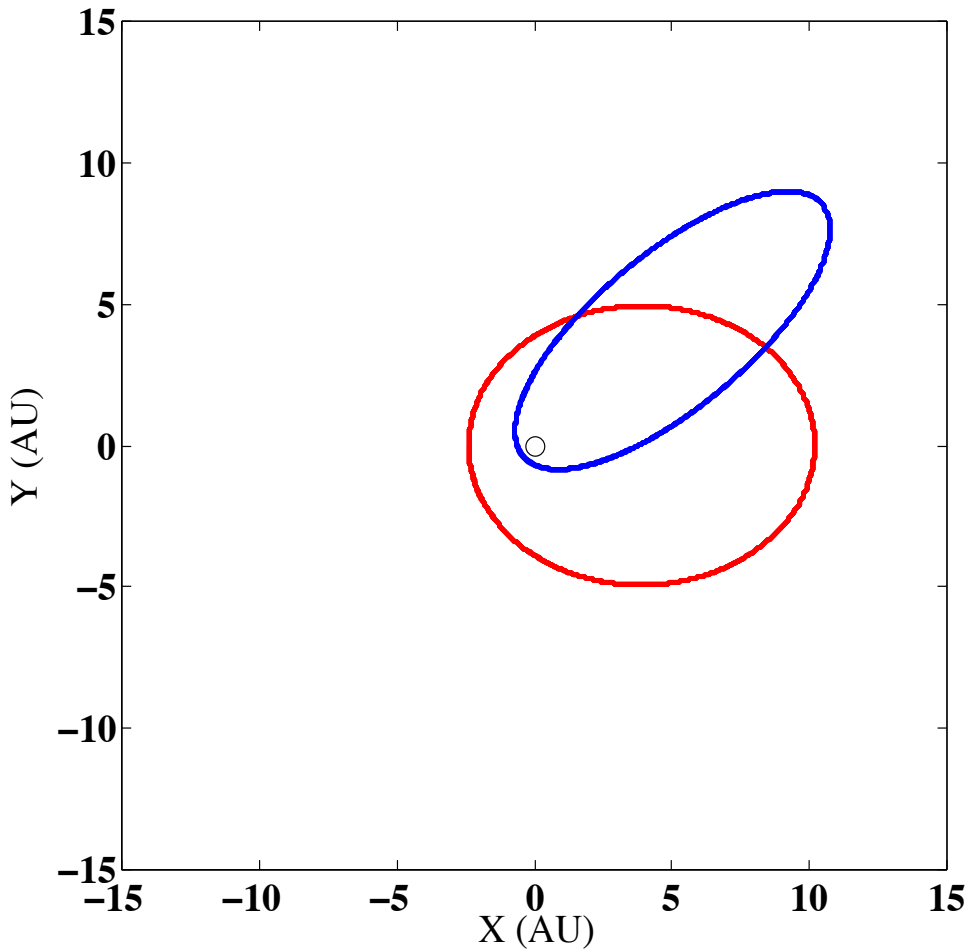

**Figure 1**: The best-fit orbits of the two proposed companions of the cataclysmic variable QS Virginis, as proposed in Almeida & Jablonski (2011). The orbit of QS Vir (AB) b is plotted in red, whilst that of QS Vir (AB) c is plotted in blue. The orbit of QS Vir (AB) c is so eccentric that it passes periastron at just 0.563 AU, whilst its apastron distance is 13.52 AU – an orbit more reminiscent of the comets in our own Solar system than the planets.

Given that the orbits of the two proposed companions are so extreme, it seems highly unlikely that they would be dynamically stable on long timescales. As such, we chose to carry out a detailed dynamical study of the proposed system, to see whether the observed timing variations could really be the result of the proposed companions.

### 3. Dynamically testing the QS Virginis System

To study the dynamical stability of the QS Virginis system, as proposed in Almeida & Jablonski (2011), we used the Hybrid integrator within the *N*-body dynamics package MERCURY (Chambers, 1999). Following our earlier work (e.g. Marshall et al., 2010; Horner et al., 2011; Horner et al., 2012a, b; Wittenmyer et al., 2012a, 2013a), we chose to hold the initial orbit of the innermost planet (QS Vir (AB) b) fixed at the nominal best-fit values given in Table 1[4]. For simplicity, and following our earlier work, we treated the central stars (the 0.78 $M_\odot$ primary and 0.43 $M_\odot$ secondary) as a single body located at the system's barycentre, with mass 1.21 $M_\odot$.

---

[4] In each of our studies, we have chosen to hold the initial orbit of one of the proposed planets fixed, and moved the other. This means that we can more thoroughly sample the orbital element space around the nominal best-fit orbits within a reasonable period of time (by choosing to vary only four variables, rather than eight!). It also allows us to avoid duplicating dynamical architectures – rotating the orbits of both planets by sixty degrees (in the same direction) would do nothing to change their dynamics, nor would scaling their orbits inwards or outwards by a given percentage. Moving the orbits of both planets would clearly duplicate many scenarios, leading to a great waste of computing time to no benefit to our study.

Since the orbital period of the stars is just over three and a half hours (compared to the ~14 year orbital period of the inner of the proposed planets), this treatment is dynamically justified[5].

We then ran a suite of 126075 simulations, within which QS Vir (AB) c started on a unique orbit, ranging across the full ± 3 sigma uncertainties in the semi-major axis (*a*), eccentricity (*e*) and argument of periastron ($\omega$) of its proposed orbit. In total, we tested 41 unique values of semi-major axis and eccentricity for QS Vir (AB) c, distributed in even steps across the full ± 3 sigma range of allowed values. For each of these 1681 *a-e* locations, we tested 15 unique values of the argument of periastron. Finally, at each of the 25215 *a-e-$\omega$* locations tested, we considered 5 unique initial mean anomalies, evenly distributed around the object's proposed orbit. For these first simulations we considered the scenario where the two planets moved on coplanar orbits – in other words, their initial mutual orbital inclination was 0°. Given the extremely small pericentre distance of QS Vir (AB) c, we chose to use a relatively short time-step for our simulations of just 10 days. As in our previous work, the Hybrid integrator changeover within Mercury was set to occur at a distance of 3 Hill radii. This ensures that the simulations run as quickly as possible (making use of the symplectic integrator within the Hybrid package) whilst the planets are widely separated, but also ensures that close encounters between the planets are accurately simulated (using the slower, but more accurate Bulirsch-Stoer integrator built in to Hybrid). For an in-depth discussion of how MERCURY handles close encounters within the Hybrid code, we direct the interested reader to Chambers (1999).

Our simulations ran for a period of 100 Myr. If one of the planets collided with the other, was ejected from the system (by reaching a barycentric distance of 20 AU), or was flung into the central stars, the time at which this happened was recorded. In figure 2, we present the mean lifetimes of the QS Virginis planetary system, as a function of the initial semi-major axis and eccentricity of QS Vir (AB) c.

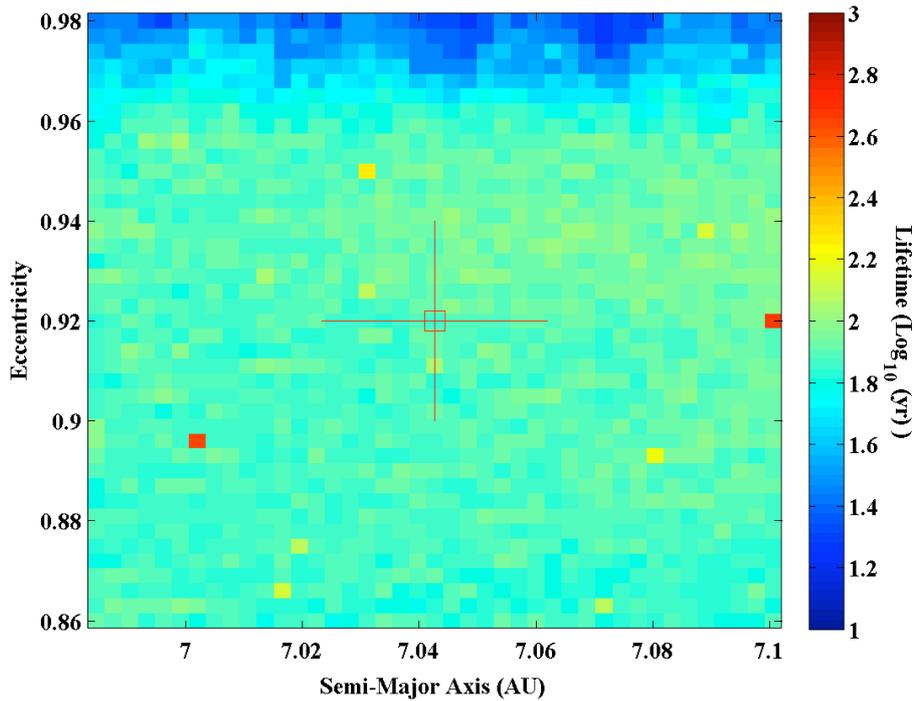

**Figure 2**: The dynamical stability of the QS Virginis planetary system, as proposed in Almeida & Jablonski, 2011, as a function of the initial semi-major axis and eccentricity of the orbit of QS Vir (AB) c. The lifetime plotted at each of the 1681 *a-e* values tested is the mean of 75 separate simulations, each of which started with QS Vir (AB) c placed on an orbit with a different combination of the argument of periastron and mean anomaly. The open wire box marks the nominal best-fit orbit for QS Vir (AB) c, whilst the lines that radiate from that location denote the 1-sigma uncertainties in that orbit. Note that even the most stable locations on the plot, which both lie well away from the nominal best-fit orbit, have mean lifetimes less than 1000 years.

---

[5] We note that the most extreme orbital solution tested for QS Vir (AB) c featured a pericentre distance of slightly less than 0.14 AU, a factor of ~25 times larger than the separation of the stars themselves. Even with such an extreme orbit, it is therefore reasonable to treat the central binary as a single object.

As can be seen in Figure 2, the proposed QS Virginis planetary system is extremely dynamically unstable, no matter what initial orbit is chosen for QS Vir (AB) c. Indeed, only two of the 126075 systems tested survived for over 8,000 years. These two "stable" outliers are the cause of the two red squares in the figure – locations where the mean lifetime is artificially increased by the presence of a single simulation with a lifetime of 28501 years (left-hand most of the red squares) and 27989 years (right-hand most). Clearly, the system is so dynamically unstable as to be unfeasible.

This instability can be illustrated by plotting the dynamical evolution of the nominal best-fit orbits for a period of 100 years, as shown in Figure 3. In stark contrast to the clearly defined orbits seen in Figure 1, it is clear that the planets interact intensely with one another from the very start of the integration – observe how neither object completes a full orbit before being perturbed by its companion.

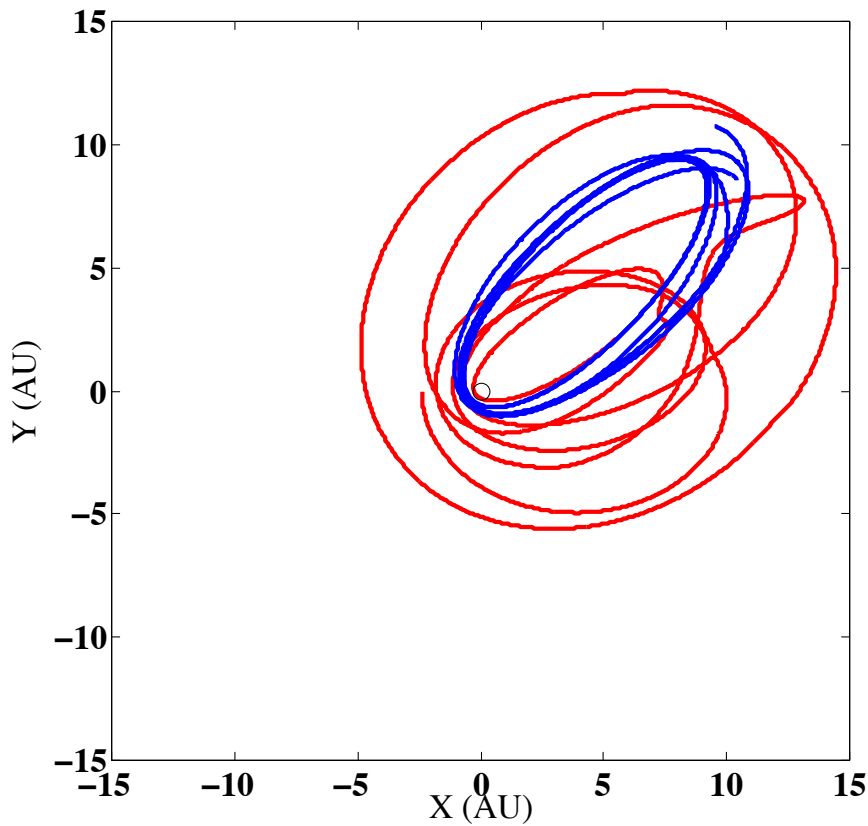

**Figure 3:** 100 years in the life of the nominal best-fit orbits proposed for the QS Virginis system in Almeida & Jablonski, 2011. As in Figure 1, the blue line shows the motion of QS Vir (AB) c, whilst the red line shows that of QS Vir (AB) b. The two planets interact so strongly that neither returns to its starting point after just one orbit.

Following our earlier work, we also examined whether the mutual orbital inclination of the two proposed planets might offer any possibility of stable orbital solutions. We ran five additional suites of integrations, at a lower resolution, featuring mutual orbital inclinations between the two planets of 5, 15, 45, 135 and 180° (as in Horner et al., 2011; Wittenmyer et al., 2012c; Wittenmyer et al., 2013a). Our results are presented in Figure 4, whilst in Figure 5, we present the same results with the colour axis ranging from lifetimes of $10^2$ to $10^8$ years, to allow direct comparison with our earlier work. In previous work, we had found that circumbinary planets moving on mutually coplanar orbits, with one moving retrograde with respect to the other (i.e. a mutual inclination of 180°) allowed some dynamically stable orbits to be found within the uncertainties on the planetary orbits (e.g. Fig. 2 of Horner et al., 2011, Fig. 2 of Wittenmyer et al., 2013a). In the case of the proposed planets around QS Virginis, however, not even this can prevent the system becoming unstable on timescales of less than a thousand years[6].

---

[6] We note, here, that to form planetary systems in which the components are highly inclined (or anti-coplanar) would clearly be challenging – and so the likelihood of such systems occurring seems low. However, given the challenges in forming/maintaining planetary systems around evolved binaries like

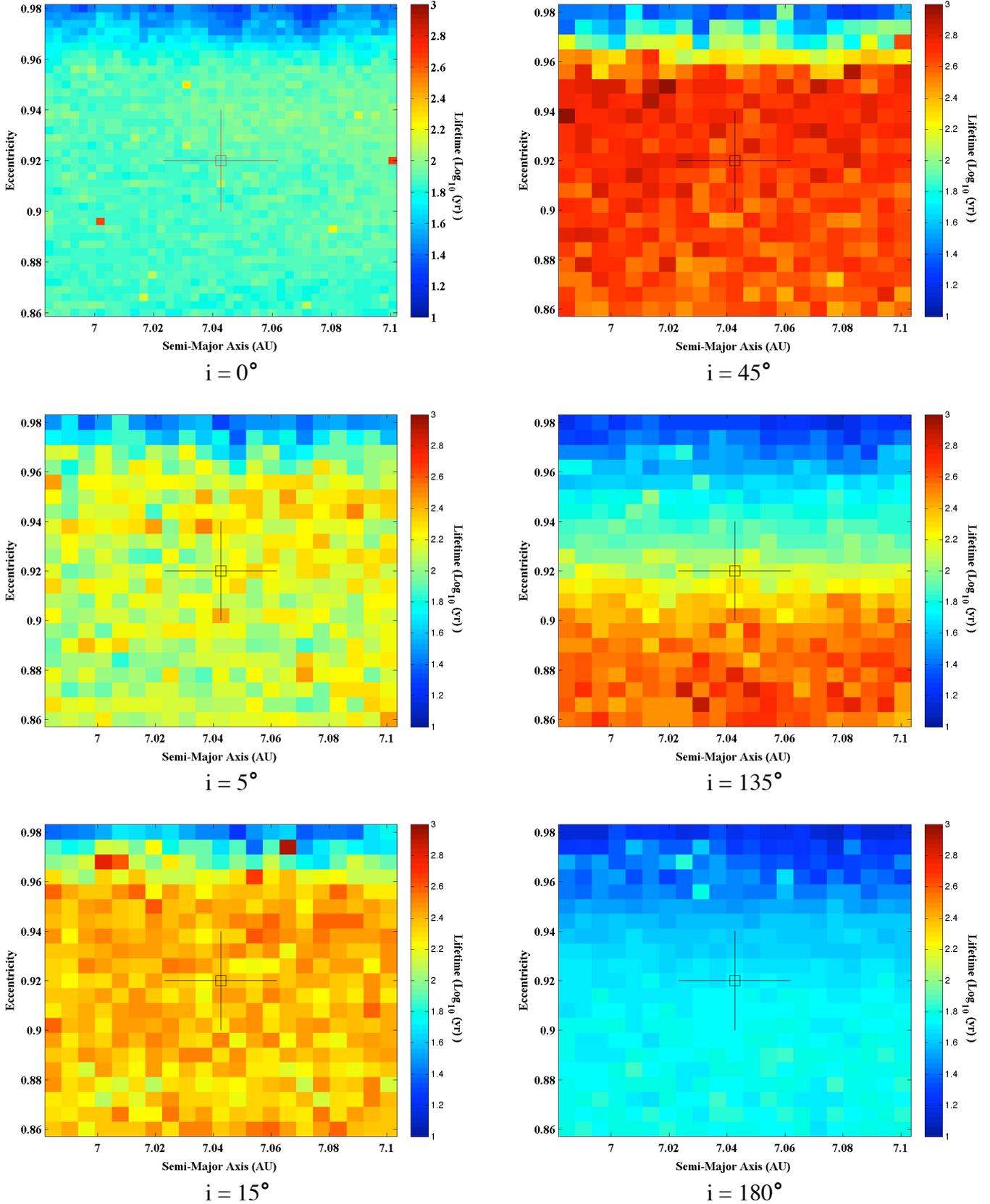

**Figure 4:** The dynamical stability of the proposed QS Virginis planetary system, as a function of the mutual orbital inclination of the two planets. The top left-hand panel shows the stability when the two

---

UZ For (e.g. Mustill et al., 2013), and given the ongoing discoveries of hot Jupiters that move on orbits that are highly inclined (or even retrograde) with respect to the spin axis of their host star (e.g. Addison et al., 2013), it is interesting to examine a range of solutions across mutual-inclination space, in case such scenarios offer prospects for unexpected orbital stability.

planets are moving co-planar orbits, while the centre-left and lower-left panels show the stability for mutual orbital inclinations of five and fifteen degrees, respectively. The upper-right panel shows the stability for a mutual inclination of 45 degrees, while the centre-right and lower-right panels show mutual inclinations of 135 and 180 degrees, respectively. As can be seen, in all cases studied, the system is dynamically unstable.

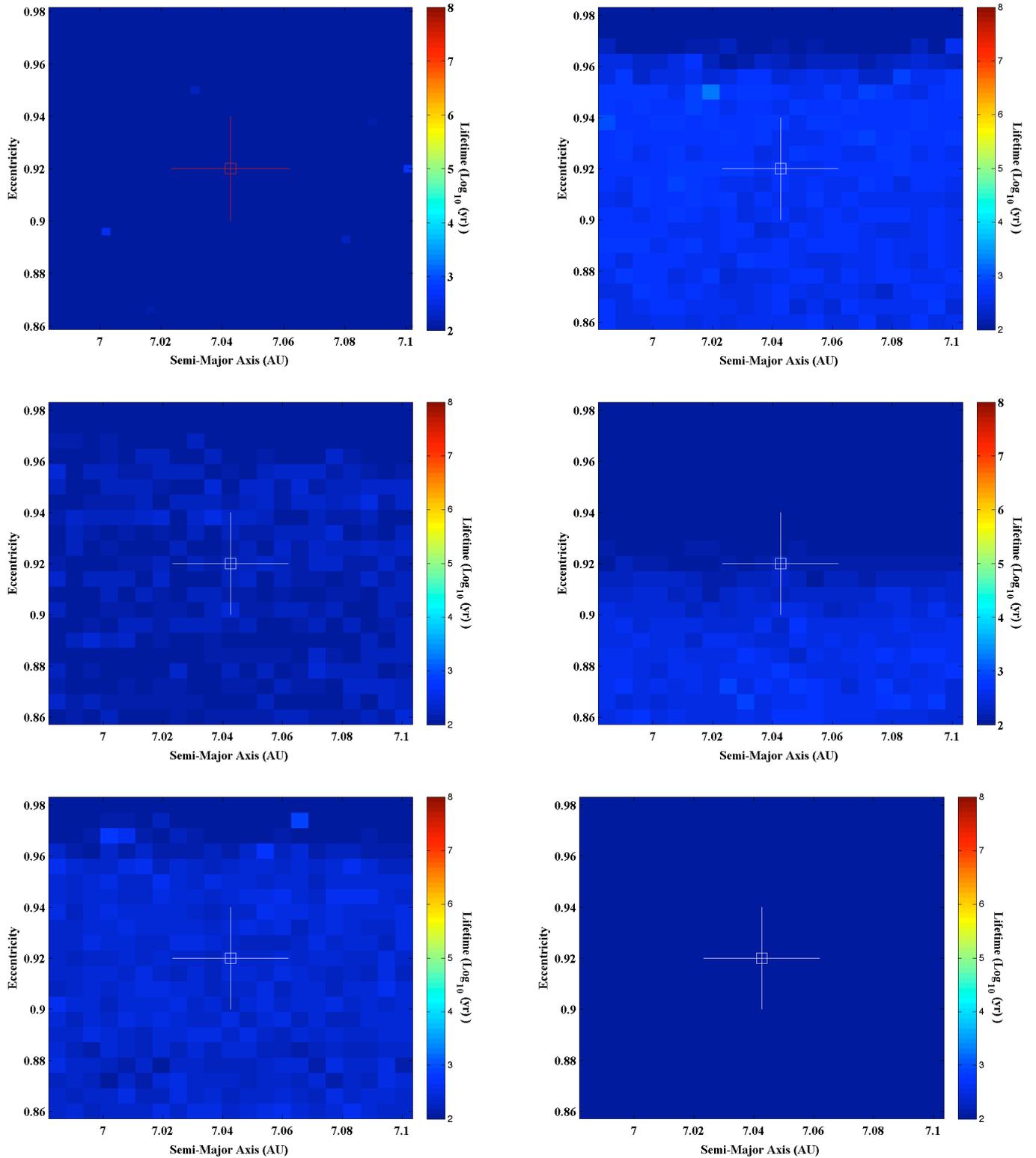

**Figure 5:** The dynamical stability of the proposed QS Virginis planetary system, as a function of the mutual orbital inclination of the planets therein. Panels as in Fig. 4. Here, the lifetimes are plotted using the same colour scheme and range of values as for our earlier work (Horner et al., 2011; Wittenmyer et al., 2013a). The extreme instability of the system is readily apparent from the sea of blue in each of the panels.

**Conclusions**

Our dynamical simulations of the proposed QS Virginis planetary system show that the proposed planets are not dynamically feasible, regardless of the mutual inclination of their orbits. In total, we carried out 181200 separate simulations of the proposed planetary system (126075 of the co-planar case, and 11025 of each of the five other mutual inclinations we considered). In all those simulations, the longest lived planetary system survived for a relatively brief 28501 years. It is fair to say that, if there are unseen companion(s) orbiting the cataclysmic variable star QS Virginis, they must move on drastically different orbits to those proposed in the discovery work (Almeida & Jablonski, 2011). Given the extreme nature of the orbits proposed for the planets in that work, this is perhaps not unsurprising. Mutually crossing orbits are almost always dynamically unstable – aside from when the objects are prevented from experiencing close encounters by the influence of mutual mean-motion resonances[7].

It seems most likely that the proposed planets do not exist, and that the observed timing variations are either the result of some other physical process (such as the chaotic nature of accretion in such systems, leading to heterogeneous data, e.g. Goździewski et al., 2012[8]), or are the result of the uncertainties in the timing of point of mid-eclipse being incorrectly determined (e.g. Hinse et al. 2013). If the timing precision has been underestimated (and the uncertainties are therefore larger than those quoted in the discovery work), then the low-amplitude signal might be within the noise level of the measurements, and would therefore be of questionable provenance. The precision with which timing uncertainties have been determined is still an unanswered question within the field of timing measurements of short-period eclipsing binaries, whose very nature makes the precise determination of eclipse mid-point a challenging process. However, the scale of the variations in O-C described by Almeida & Jablonski (2011) are so large (up to 200s) that their uncertainties may well not be the only cause of the observed deviation from a linear ephemeris.

Future observations of the system should help to resolve this issue once and for all – if the observed timing variations are simply the result of poorly estimated uncertainties in the timing of the eclipses, then they will likely disappear entirely in the coming years. If the variations continue to be observed, however, then an interesting question is posed – what is causing the eclipses between the components of the QS Virginis system to vary in time?

**Acknowledgements**
The authors wish to thank the referee, Ramon Brasser, for providing swift and helpful feedback. The work was supported by iVEC through the use of advanced computing resources located at the Murdoch University, in Western Australia. This research has made use of NASA's Astrophysics Data System (ADS), TCH gratefully acknowledges financial support from the Korea Research Council for Fundamental Science and Technology (KRCF) through the Young Research Scientist Fellowship Program, and also the support of the Korea Astronomy and Space Science Institute (KASI) grant 2013-9-400-00. JPM is supported by Spanish grant AYA 2011/26202. AJM is supported by Spanish grant AYA 2010/20630.

---

[7] Such protective resonant behavior is widely seen in our own Solar system, with the Plutino population trapped in 3:2 resonance with Neptune (e.g. Malhotra, 1995; Chiang & Jordan, 2002) and the planetary Trojans, trapped in 1:1 resonance with their host planet (e.g. Jewitt et al., 2000; Sheppard & Trujilo, 2006; Horner & Lykawka, 2010).

[8] Given the dispute over the nature of the QS Virginis (e.g. Parsons et al., 2011), we note that the chaotic influence of accretion would obviously only by important for currently accreting CVs, and not for those that are hibernating or detached. If QS Virginis falls into the latter two classes, then this mechanism can clearly not be invoked in this case.